\documentclass[sigconf,nonacm,conference]{acmart}
\renewcommand\footnotetextcopyrightpermission[1]{} 
\pdfoutput=1
\usepackage[noend]{algorithmic}
\usepackage{algorithm}
\usepackage{listings}
\usepackage{xcolor}
\usepackage{tabularx}
\usepackage{arydshln}
\usepackage[inkscapelatex=false]{svg}
\usepackage{fancyhdr}
\definecolor{mygray}{rgb}{0.5,0.5,0.5}
\lstdefinestyle{mystyle}{
  backgroundcolor=\color{white},
  commentstyle=\color{mygray},
  keywordstyle=\color{blue},
  numberstyle=\tiny\color{mygray},
  stringstyle=\color{red},
  basicstyle=\ttfamily\footnotesize,
  breakatwhitespace=false,
  breaklines=true,
  captionpos=b,
  keepspaces=true,
  numbers=none,
  numbersep=5pt,
  showspaces=false,
  showstringspaces=false,
  showtabs=false,
  tabsize=2,
  xleftmargin=1em,  
  xrightmargin=1em,  
  belowskip=0.5em,  
  aboveskip=0.5em,  
}
\newcommand{\sysname}{Moctopus}
\AtBeginDocument{%
  }

\begin{document}

\title{Accelerating Regular Path Queries over Graph Database with Processing-in-Memory}

\author{Ruoyan Ma\textsuperscript{1}, Shengan Zheng\textsuperscript{1}, Guifeng Wang\textsuperscript{1}, Jin Pu\textsuperscript{1}, Yifan Hua\textsuperscript{1}, Wentao Wang\textsuperscript{2}, Linpeng Huang\textsuperscript{1}}
\thanks{%
  Shengan Zheng (shengan@sjtu.edu.cn) and Linpeng Huang (lphuang@sjtu.edu.cn) are corresponding authors. 
  Shengan Zheng is with MoE Key Lab of Artificial Intelligence, AI Institute, Shanghai Jiao Tong University.
  This work is supported 
  by National Key Research and Development Program of China (Grant No. 2022YFB4500303), 
  National Natural Science Foundation of China (NSFC) (Grant No. 62227809), 
  the Fundamental Research Funds for the Central Universities, and
  Shanghai Municipal Science and Technology Major Project (Grant No. 2021SHZDZX0102).
}
\affiliation{%
  \institution{\textit{\textsuperscript{1}Shanghai Jiao Tong University}\quad\textit{\textsuperscript{2}Peking University}}
  \country{}}

\begin{abstract}
  Regular path queries (RPQs) in graph databases are bottlenecked
  by the memory wall. 
  Emerging processing-in-memory (PIM) technologies offer a promising solution to dispatch 
  and execute path matching tasks in parallel within PIM modules.
  We present \sysname{}, a PIM-based data management system for graph databases that supports efficient batch RPQs and graph updates.
  \sysname{} employs a PIM-friendly dynamic graph partitioning algorithm, 
  which tackles graph skewness and preserves graph locality with low overhead for RPQ processing.
  \sysname{} enables efficient graph update by amortizing the host CPU's update overhead to PIM modules.
  Evaluation of \sysname{} demonstrates superiority over the state-of-the-art traditional graph database.
\end{abstract}



\maketitle

\section{Introduction}
The rise of graph data in volume and complexity has spurred a growing interest
in graph databases from both academic\cite{chen2022g, martens2022representing, besta2023graph} 
and industries\cite{tian2023world,li2022bytegraph}. 
\emph{Regular path queries} (RPQs) are one of the most essential classes of queries on graph databases.
When evaluating RPQ over a graph, graph databases return all endpoint pairs of
matched paths in the graph.

Unfortunately, RPQs on traditional graph databases face the "memory wall"
bottleneck. Processing RPQ involves a large amount of pointer chasing, which
triggers a considerable number of random memory accesses for accessing the
neighborhood nodes, forcing most data accesses to use DRAM memory rather than
cache. The excessive data movement results in high access latency and high bandwidth consumption 
that constrain the performance and incur considerable energy costs,
which is also known as the "memory wall".

Processing-in-Memory (PIM) \cite{mutlu2022modern,upmem} enables computations
and processing within the device's memory, which offers a promising solution to the "memory wall" challenge.
PIM systems are typically structured with a powerful host CPU and a set of PIM
modules with wimpy cores. The host CPU dispatches data-intensive computational
tasks to PIM modules, and gathers the results after the PIM modules finish the computation. The PIM modules execute these
data-intensive computations within the memory modules that integrate
computational resources, reducing data movement to the host CPU for processing.
This results in improved performance for applications that deal with high data
intensity or suffer from low cache locality
\cite{giannoula2021syncron,gomez2021benchmarking}.
PIM has been widely used in the field of graphs and databases for graph analysis
\cite{ahn2015scalable,huang2020heterogeneous,zhang2018graphp} and database
indexing\cite{kang2023pim,kang2023pim2}.




PIM systems face significant challenges for performing RPQs over graph
databases. The first challenge is the imbalanced load distribution among PIM
modules, especially with highly skewed graphs. Many real-world graphs
\cite{faloutsos1999power,papalexakis2016power,newman2005power} exhibit varying
degrees of skewness, characterized by few high-degree (out-degree) nodes and
many low-degree nodes. High-degree nodes need more computing and bandwidth
resources given their large neighborhood. Consequently, PIM systems often suffer
from a load imbalance situation, where some PIM modules are overloaded with
high-degree nodes, while others are underutilized. The second challenge is the
high communication overhead of PIM-based graph databases, which consists of
CPU-PIM communication (CPC) and inter-PIM communication (IPC). On the commodity
UPMEM\cite{upmem} platform, the bandwidth of these two communication modes is
less than 2\% of intra-PIM bandwidth\cite{gomez2021benchmarking}. To minimize
IPC overhead, the graph partitioning algorithm needs to preserve graph
locality, as many next-hops in different PIM modules would incur high IPC costs
otherwise. The third challenge is dynamic graph management with constant
insertion and deletion of nodes and edges, and the graph storage engine needs
to efficiently handle frequent updates.

We present \sysname{}, a PIM-based data management system for graph databases that
leverages the unique features of PIM to achieve high performance for RPQs and
graph updates. By harnessing the parallel capabilities of PIM modules,
\sysname{} significantly accelerates path matching and graph update operations.
\sysname{} relies on a novel graph partitioning algorithm that exploits the PIM features
with \emph{locality-aware node distribution} and \emph{greedy-adaptive load balancing}.
To handle different types of workloads, \sysname{} adopts a labor-division approach 
that leverages the strengths of both the host CPU and the PIM modules.
Specifically, high-degree nodes demonstrating a good locality access pattern are
assigned to the host CPU, whereas low-degree nodes are assigned to PIM modules.
Thus, the PIM modules can overcome the load imbalance issue 
that stems from graph skewness by avoiding the high-degree nodes.
Instead of randomly assigning graph nodes to PIM modules using a hash function, we can record the partitioning state 
by keeping track of previous partitioning decisions and achieve more precise graph partitioning among PIM modules. 
To preserve graph locality among PIM modules with low overhead, we
propose a greedy-adaptive method that combines the greedy method and adaptive method. 
The two-stage load balancing method uses a radical greedy heuristic to reduce
partitioning overhead, then migrates incorrectly partitioned nodes to enhance graph locality.
Besides, it uses a dynamic capacity constraint to enforce load balance across PIM modules.
The graph partitioning algorithm achieves PIM-friendliness 
by balancing the workload and preserving the locality of the graph among PIM modules.
For graph updates, high-degree nodes are more likely to be updated frequently and consume more CPU resources. 
To address this challenge, \sysname{} employs heterogeneous graph storage for high-degree nodes to ease the CPU's load
and delegate complex update operations to PIM modules.

We implement \sysname{} on the commodity PIM system, UPMEM\cite{upmem}. We
compare \sysname{} with RedisGraph\cite{RedisGraph} and the scheme of 
the widely-used hash using real-world graphs to demonstrate \sysname{}'s
superiority. RedisGraph is a state-of-the-art in-memory graph database.
The benchmark of processing a typical RPQ, k-hop path query on 15 real-world
graphs shows that \sysname{} is up to 10.67x faster than RedisGraph. Compared
with the widely-used hash scheme, \sysname{} effectively reduces communication overhead. 
Furthermore, fully exploiting high parallel intra-PIM bandwidth, \sysname{} is remarkably faster than RedisGraph
regarding graph update, with an average of 30.01x for insertion and 52.59x for deletion.

In summary, this paper makes the following contributions:

(1) We present \sysname{}, a PIM-based data management system for graph databases 
that supports efficient batch RPQs and graph updates.
To the best of our knowledge, it's the first design to
accelerate path matching queries over graph database with PIM.

(2) We propose a PIM-friendly dynamic graph partitioning algorithm that
tackles graph skewness and preserves graph locality with low overhead.
With PIM-friendly graph partitioning, \sysname{} successfully addresses
the challenges of load imbalance and communication bottleneck during performing RPQs.

(3) We achieve efficient graph update with heterogeneous graph storage for high-degree nodes
by amortizing the host side's update cost to the PIM side.

(4) We implement and evaluate the \sysname{} on a commercial PIM system,
demonstrating the \sysname{}'s superiority.
Compared with the state-of-the-art traditional graph database, 
our system achieves up to 10.67x speedups for RPQ and 209.31x speedups for graph update.


\section{background}

\subsection{Graph Database and Graph Partition}

As the volume of data continues to expand, traditional single-node graph databases are no
longer adequate to meet the growing demands. More databases seek to partition
graphs across multiple computing nodes. The PIM architecture is similar to the
distributed graph database, for both involve multiple nodes participating in
computation.

\textbf{Graph database.}
Graph databases utilize the property graph model\cite{angles2018property} to
represent graph data. In a property graph, nodes represent
distinct entities, and (directed) edges are employed to depict the
relationships between pairs of entities. Nodes and edges have labels
and property-value pairs to describe their attributes. For processing RPQs,
paths composed of entities and relationships are considered first-class
citizens. Focusing on path matching and for simplicity, we use an adjacency
matrix to present a simplified property graph (directed graph), in which non-essential features (labels and
property-value records) are excluded. 

\textbf{Graph partition for distributed graph database.}
There are two graph partition solutions for the existing distributed graph database:

(1) Master-slave replication.
Master-slave replication refers to replicating data from a primary database (master)
to one or more secondary databases (slaves).
The most popular graph database, Neo4j\cite{Neo4J}, adopts this solution.
For this approach, each computing node stores the global graph.
Nevertheless, the PIM module is constrained by limited local memory capacity (for UPMEM, 64MB),
which renders the storage of the complete global graph nearly unfeasible.

(2) Hash partition.
This widely-used method assigns graph nodes to computing nodes according to a consistent hashing function.
The representative distributed graph databases are G-Tran \cite{chen2022g} and ByteGraph \cite{li2022bytegraph}.
However, since graph nodes are randomly assigned to PIM modules,
this method overlooks the locality within the graph, resulting in high IPC overhead.
Besides, graph skewness causes severe load imbalance among PIM modules.

\textbf{Graph partition for preserving graph locality.}
There are two graph partition solutions for preserving graph locality: the greedy method and adaptive method.
Linear Deterministic Greedy (LDG) \cite{stanton2012streaming} is a good representative of the geedy method.
LDG uses a greedy heuristic that assigns a graph node to the partition containing most of its neighbors,
effectively preserving graph locality.
However, the assignment could be more time-consuming for traversing all computing nodes.
Besides, LDG is unsuitable for dynamic graphs within graph databases
because it requires prior knowledge of the graph structure, such as the number of nodes and edges.
For the adaptive method\cite{vaquero2013adaptive}, new graph nodes are randomly
assigned to computing nodes according to a hash function, and computing nodes
iteratively migrate graph nodes to generate partitions with good locality. This
technique supports dynamic graphs with nodes and edges constantly inserted and
deleted but has a huge communication overhead for migrating nodes.

The graph partitioning algorithm in \sysname{} leverages the strengths of both
greedy method and adaptive method.

\vspace{-2.mm}

\subsection{PIM architecture}

With the UPMEM, Processing-in Memory DIMMs are already commercially available.
The PIM model comprises two main components: a powerful host CPU with supreme
controlling authority (the host side) and a set of $P$ PIM modules (the PIM
side). Each PIM module consists of an on-bank processor (PIM processor) and a
local memory (for UPMEM, 64MB). Despite its simplicity, the PIM processor
remains general-purpose. It facilitates a range of interactions between the
host CPU and the PIM modules, allowing the host CPU to transmit executable code
to the PIM modules, launch the code, and monitor its completion. IPC can be
realized by leveraging CPC through CPU forwarding data. The bandwidth of CPC
and IPC is expensive. 
Despite having 2048 PIM modules and delivering 1.28TB/s of intra-PIM bandwidth\cite{gomez2021benchmarking}, 
the system can only offer roughly 25GB/s total CPC and IPC bandwidth.

\subsection{Matrix-based graph operations}

Graphs can be represented by matrices, and graph algorithms can be implemented
by matrix-based operations. In graph database scenarios, graph pattern matching
can be translated into a set of matrix multiplications. The
GraphBLAS\cite{kepner2015graphs}, an established mathematical framework,
defines a core set of matrix-based graph operations that can be used to
implement a wide class of graph algorithms. Owing to the high serial and parallel
processing performance of matrix-based operations, RedisGraph employs GraphBLAS
to support efficient graph queries\cite{RedisGraph}.

Similar to RedisGraph, \sysname{}'s execution plan is composed of matrix-based
graph operations, aiming to exploit the parallelism of PIM modules by
leveraging the natural parallelism of matrix operations. For example, if we
want to find two hops away nodes from fixed source nodes (the batch 2-hop path
query in Figure~\ref{fig:graphPartitionExample}), the matrix-based execution
plan would be $ans = Q \times Adj \times Adj$, where $Q$ is a matrix containing
source nodes information, $Adj$ stands for the adjacent matrix, and $ans$ is a
matrix containing destination nodes information. The rows of the $Q$ identify a
query in a batch of queries, and the columns represent source nodes. The rows
of the $ans$ identify a query in a batch of queries, and the columns represent
destination nodes.

\section{design}

In this section, we introduce the design of the \sysname{}, supporting
efficient batch RPQs and graph updates. The key design is a PIM-friendly graph
partitioning algorithm, aiming to reduce communication costs and
achieve load balance during path matching.

\subsection{System architecture}

\begin{figure}[t]
  \centering
  \includegraphics[width=7.5cm]{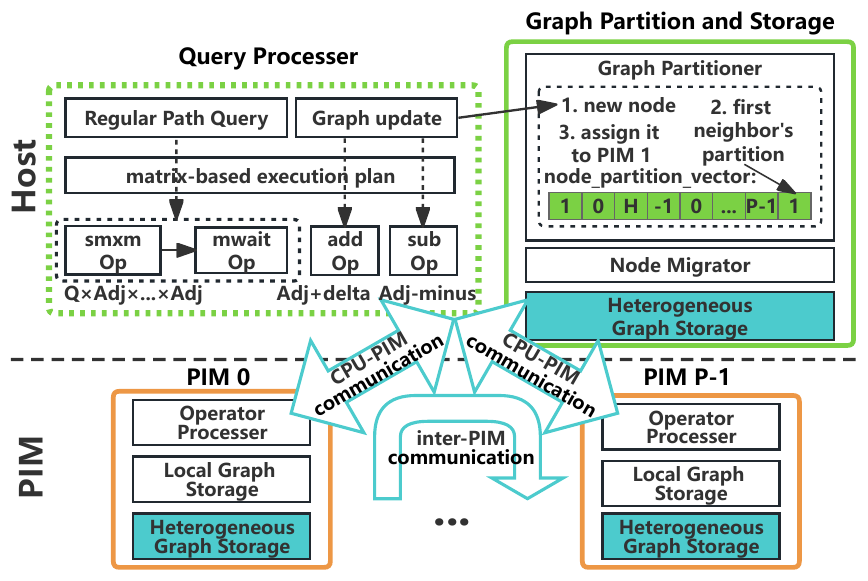}
  \vspace{-4.mm}
  \caption{The architecture of the \sysname{} with $P$ PIM modules.}
  \vspace{-6.mm}
  \label{fig:architecture}
\end{figure}

Figure~\ref{fig:architecture} shows the overall design of the \sysname{}.
\sysname{} adopts an adjacency matrix to represent a graph and uses matrix-based
operations to query the graph. The graph is partitioned across the host side and PIM side.
Leveraging both sets of resources, \sysname{} achieves efficient RPQs and graph updates.
As depicted in Figure~\ref{fig:architecture}, the main components of \sysname{}
include:

1) The Query Processor.
When processing batch RPQs and graph updates, the Query Processor generates execution plans composed of matrix-based operators,
and dispatches these operators to PIM modules for processing.
RPQ will be translated into a \texttt{smxm} operator for path matching and a \texttt{mwait} operator for reducing the result. 
Graph update is abstracted into \texttt{add} operator and \texttt{sub} operator.
Like the map-reduce model, inherently parallel tasks (matrix-based operators) are mapped to $P$ PIM modules for execution,
exploiting the high parallel intra-PIM bandwidth.


2) The Graph Partitioner and Node Migrator.
To fully leverage the potential of PIM and achieve efficient path matching, the graph
should be partitioned across computing nodes (the host CPU and $P$ PIM modules).
Through the collaboration of the Graph Partitioner and Node Migrator,
\sysname{} achieves dynamic graph partitioning.
The graph partitioning algorithm performs a disjoint partitioning by graph node
among computing nodes. It is PIM-friendly, achieving load balance among PIM modules and maintaining
graph locality with low overhead.
The Graph Partitioner assigns new graph nodes according to a radical greedy heuristic (described in Section 3.2.2).
As shown in Figure~\ref{fig:architecture}, when processing graph updates,
if an endpoint node appears for the first time in the inserting edge stream,
the Graph Partitioner will identify it as a new node and assign it 
considering the history partitioning decisions stored in \texttt{node\_partitioning\_vector}.
The Node Migrator is responsible for relocating new high-degree nodes to the host side
and migrating incorrectly partitioned nodes to appropriate partitions.

3) The Operator Processor.
Each PIM module contains an Operator Processor.
The Operator Processor is responsible for parsing and processing operators
received from the host CPU.

\begin{figure}[t]
  \centering
  \includegraphics[width=7.5cm]{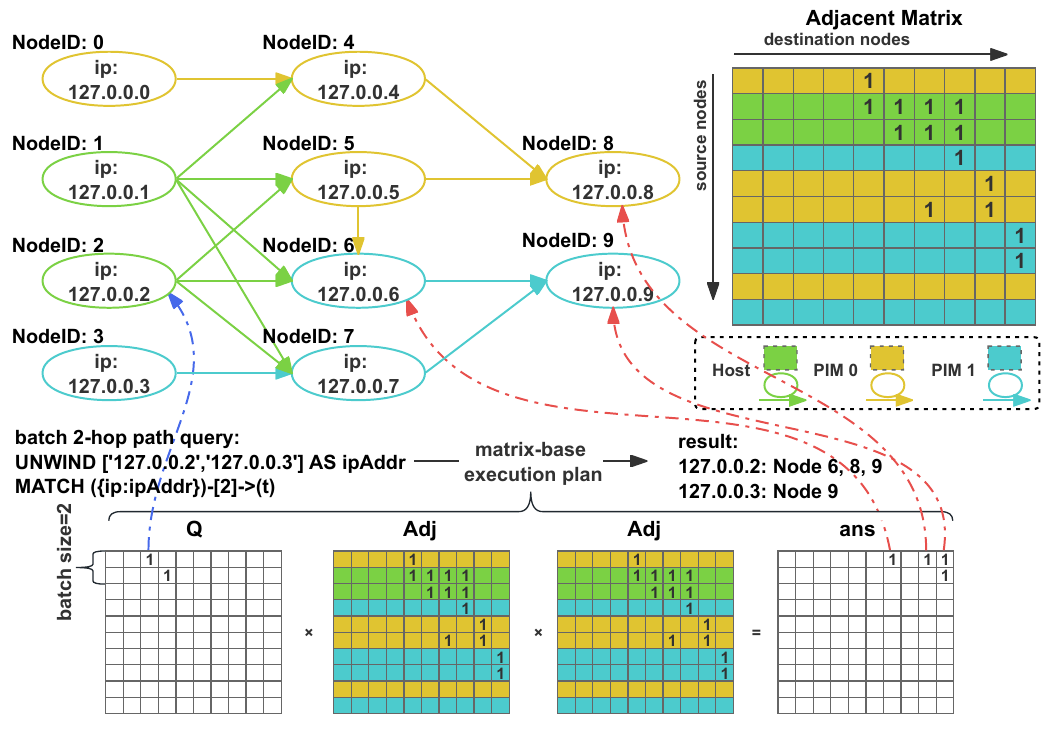}
  \vspace{-4.mm}
  \caption{Example of partitioning a routing connection graph in property graph and adjacency matrix view,
    and the matrix-based execution plan of a batch 2-hop path query.}
  \vspace{-6.mm}
  \label{fig:graphPartitionExample}
\end{figure}

4) The Local Graph Storage and Heterogeneous Graph Storage.
Since \sysname{} adopts a disjoint partitioning by graph node,
as shown in Figure~\ref{fig:graphPartitionExample}, the adjacency matrix is partitioned across the $1+P$
computing nodes by row, and every computing node maintains an adjacency matrix segment.
Given hash map's excellent concurrency and scalability,
each PIM module stores corresponding adjacency matrix segment in local graph storage using a hash map.
The hash map stores the mapping of row ID (NodeID) to row data (the next-hop data, i.e., NodeIDs of next-hop).
Nevertheless, simple local graph storage cannot meet the demands of frequent updates on the host side.
By introducing heterogeneous graph storage, we make optimizations for 
storing the adjacency matrix segment maintained by the host CPU.


In the remainder of this section, we will describe in detail the PIM-friendly dynamic
graph partitioning algorithm (Section 3.2) and optimizations for graph storage
(Section 3.3).

\vspace{-2.mm}

\subsection{Graph partition}

\sysname{} takes graph partition as the first-order design consideration, aiming to
achieve load balance among PIM modules and preserve graph locality with low overhead.
In this section, we describe the PIM-friendly graph partitioning algorithm
composed of a labor-division approach and a greedy-adaptive method.
For the labor-division approach, \sysname{} treats high- and low-degree nodes differently
to tackle graph skewness and leverage the strengths of both the host CPU and the PIM modules.
For the greedy-adaptive method, \sysname{} tries to assign and migrate adjacent nodes to the same PIM module under low overhead,
aiming to reduce IPC overhead during path matching.

\subsubsection{locality-aware node distribution}

To address load imbalance caused by graph skewness and leverage the strengths
of the host side and the PIM side, we propose a labor-division approach that
the host side handles high-degree nodes and the PIM side deals with
low-degree nodes. As shown in Figure~\ref{fig:graphPartitionExample}, the host
CPU handles nodes 1 and 2 (high-degree nodes), and the rest nodes are
partitioned into PIM 0 and PIM 1.


\textbf{Migrate high-degree nodes to the host side.}
High-degree nodes tend to be accessed more frequently and need more computing
and bandwidth resources, causing serve load imbalance among PIM modules. As the
graph grows, a low-degree node may have more connections and turn into a high-degree node. At this point,
the Node Migrator will migrate the node from the PIM side to the host side.
Since PIM modules no longer handle high-degree nodes, the load imbalance caused
by graph skewness naturally dissipates.

\textbf{Leverage the strengths of the host side and the PIM side.}
The labor-division approach also concurrently aligns with the PIM architecture.
The two components of the PIM architecture, the host side and the PIM side, prefer different workloads.
The distributed PIM side prefers uniformly random memory access workloads to exploit high parallel intra-PIM bandwidth,
while the host side with a powerful CPU prefers continuous and skewed memory access workloads with good locality.
Correspondingly, high-degree nodes tend to be accessed more frequently, and
fetching their large amounts of next-hop's NodeIDs exhibits continuous memory
access, satisfying the host side's preference. Low-degree nodes with few next
hops are likely to be accessed randomly, satisfying the PIM side's taste.
Through dispatching path matching tasks associated with high-degree nodes to
the CPU side and low-degree nodes to the PIM side, the \sysname{} can leverage
the advantages of both sets of resources.


\subsubsection{greedy-adaptive load balancing}

To preserve graph locality with low overhead and support dynamic graphs,
we propose a greedy-adaptive method to partition low-degree nodes among PIM modules.
The greedy-adaptive method combines the greedy method and adaptive method.
When receiving a new graph node, the method uses a radical greedy
heuristic that assigns it according to its first neighbor.
The radical greedy heuristic significantly reduces partitioning overhead but might cause incorrectly
partitioned nodes. As performing path matching, PIM modules detect these
incorrectly partitioned nodes, then the host CPU migrates them to correct
partitions to enhance graph locality. 
\textbf{Balance locality and overhead.}
The key to preserving graph locality is placing adjacent nodes on the same PIM module.
The radical greedy heuristic does not aim for optimal locality, 
but rather a satisfactory trade-off between graph locality and partitioning overhead. 
Instead of assigning a graph node to the partition containing most of its neighbors, the radical
greedy heuristic employs a more assertive approach by assigning a graph node to
the partition housing its first neighbor.
For PIM systems, the former method requires traversing numerous PIM modules, potentially up to tens or hundreds, 
in order to determine the appropriate partition, leading to substantial partitioning overhead. 
The radical greedy heuristic sacrifices some graph locality and tolerates a few incorrectly partitioned, 
but it only requires minimal partitioning overhead.

\textbf{Enhance locality by migration.}
When performing RPQs, \sysname{} utilizes an adaptive method to recover the
graph locality sacrificed for low partitioning overhead. During path matching,
PIM modules simultaneously detect incorrectly partitioned nodes that miss most of the next-hop
nodes in the local PIM module, effectively overlapping detection overhead with
path matching query processing. Then, the host CPU migrates them to the partitions
containing most of their neighbors. With more neighbors on the same module,
\sysname{} enhances graph locality.

It is worth noting that the labor-division
approach (Section 3.2.1) makes a graph easier to partition on the PIM side, as high-degree nodes 
have been migrated to the host side.
Ideally, the graph without high-degree nodes becomes multiple
disconnected subgraphs. In most cases, the radical greedy method
has well maintained graph locality among PIM modules, only leaving a few nodes
to migrate. Ultimately, the CPU only has to deal with a small amount of migration overhead.

\textbf{Radical greedy heuristic and migration brings flexibility.}
Besides its low partitioning overhead, a critical factor in our decision to employ the radical greedy heuristic
is its flexibility.
\sysname{} makes graph node assignment decisions upon inserting the first edge of a graph node
rather than delaying until the whole graph is established, 
accommodating the demands of partitioning dynamic graphs in graph databases.
Additionally, migration adapts to graph changes.
As the graph changes over time, the partitioning accuracy may deteriorate, 
leaving some incorrectly partitioned nodes.
\sysname{} migrates these nodes at runtime to maintain graph locality as the graph evolves.


\textbf{Enforce load balance by a dynamic constraint.}
When assigning graph nodes, we use a dynamic assigned node's capacity constraint to enforce load balance across PIM modules.
The dynamic capacity constraint is set to 1.05x the average number of assigned nodes among PIM modules,
increasing with graph scale.
When the number of assigned nodes in a PIM module exceeds the capacity constraint,
new graph nodes will be allocated into the PIM modules
below the capacity constraint using a hash algorithm
to avoid the load imbalance case where most graph nodes are assigned to a few PIM modules.
Decreasing the proportion of capacity constraint can facilitate load balance
but at the expense of decreased graph locality.


Eventually, the PIM-friendly graph partitioning algorithm achieves load balance
among PIM modules and maintains graph locality with low overhead.
Additionally, it is flexible enough to support dynamic graphs in graph
databases. 

\begin{figure}[t]
  \centering
  \includegraphics[width=8cm]{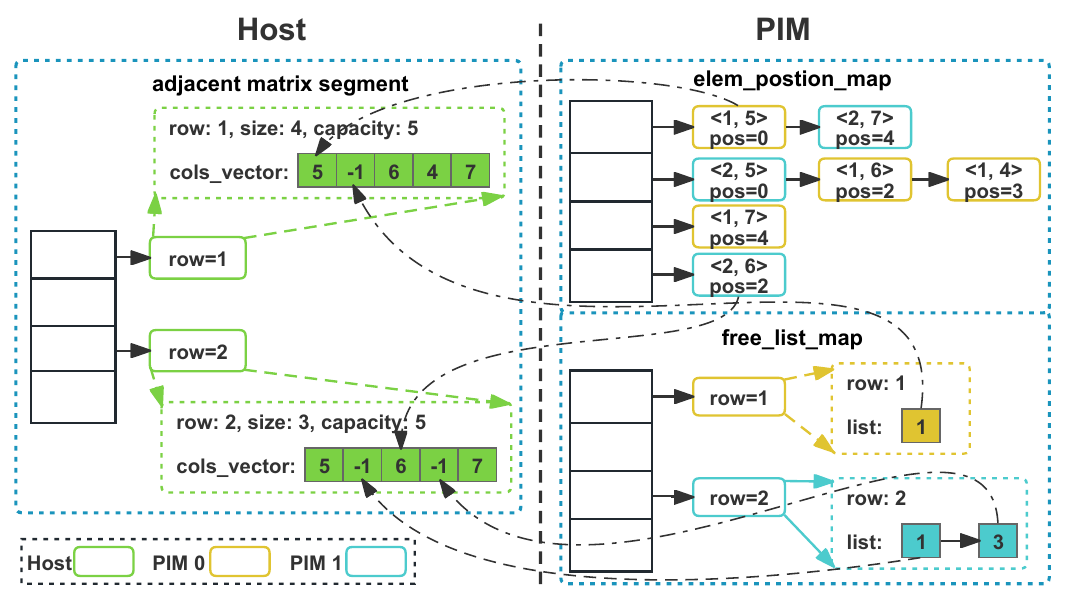}
  \vspace{-4.mm}
  \caption{Heterogeneous graph storage for high-degree nodes.}
  \vspace{-6.mm}
  \label{fig:GraphStorageOptimizations}
\end{figure}


\subsection{Optimizations for graph storage}
In \sysname{}, every computing node needs to manage graph nodes assigned to it
and maintains an adjacent matrix segment to store these graph data. PIM modules
can leverage high parallel intra-PIM bandwidth to provide high performance for querying and
updating low-degree nodes. Nevertheless, high-degree nodes on the host side
tend to be accessed and updated more frequently, placing substantial pressure
on the host CPU. To alleviate the host CPU's load, \sysname{} uses
heterogeneous graph storage for high-degree nodes, achieving efficient query
and update simultaneously by amortizing the host side's update cost to the PIM
side.

\textbf{Efficient graph query.}
On the host side, the most efficient approach for graph querying is to store the next-hop data (NodeIDs of next-hop) of high-degree nodes
in a contiguous memory array (\texttt{cols\_vector} in Figure~\ref{fig:GraphStorageOptimizations}).
Consequently, when accessing a graph node, \sysname{} requires only one memory fetch to acquire its next-hop data,
further improving memory accessing locality on the host side.

\begin{table*}[ht]
  \caption{The real-world graphs from SNAP dataset used in our experiments.}
  \vspace{-5.mm}
  \begin{center}
    \resizebox{\textwidth}{!}{%
      \begin{tabular}{lccccccccccccccc}
        \toprule
        \textbf{Name}                & roadNet-CA & roadNet-PA & roadNet-TX & cit-patents & com-youtube & com-DBLP & com-amazon & wiki-Talk & email-EuAll & web-Google & web-NotreDame & web-Stanford & amazon0312 & amazon0505 & amazon0601 \\
        \midrule
        \textbf{Trace ID}            & \#1        & \#2        & \#3        & \#4         & \#5         & \#6      & \#7        & \#8       & \#9         & \#10       & \#11          & \#12         & \#13       & \#14       & \#15       \\
        \textbf{nodes}               & 1,965,206  & 1,088,092  & 1,379,917  & 3,774,768   & 1,134,890   & 317,080  & 334,863    & 2,394,385 & 265,214     & 875,713    & 325,729       & 281,903      & 262,111    & 410,236    & 403,394    \\
        \textbf{high-degree nodes\%} & 0          & 0          & 0          & 2.83        & 2.07        & 3.10     & 0.62       & 0.50      & 0.29        & 1.29       & 2.86          & 4.84         & 0          & 0          & 0          \\
        \bottomrule
      \end{tabular}%
    }
    \label{tab:dataset}
    \vspace{-4.mm}
  \end{center}
\end{table*}

\begin{figure*}[t]
  \centering
  \includegraphics[width=\linewidth]{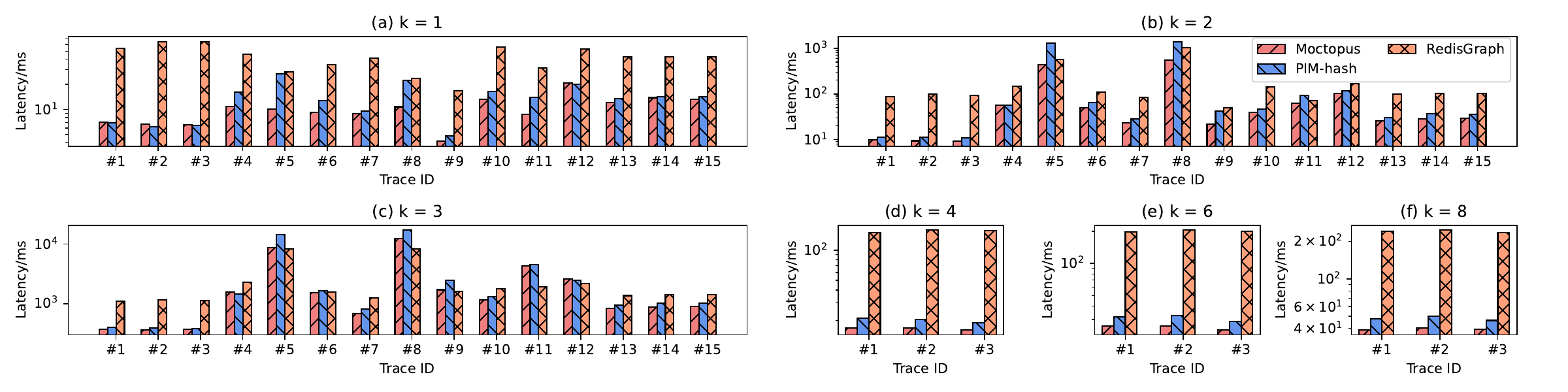}
  \vspace{-9.mm}
  \caption{Run-time of $k$-hop path queries (log scale).}
  \label{fig:kHopPathQuery}
  \vspace{-4.mm}
\end{figure*}

\textbf{Efficient graph update.}
When inserting or deleting an edge, \sysname{} has to traverse the \texttt{cols\_vector} to determine if the edge already exists.
To avoid the time-consuming traversal,
\sysname{} maintains two supplementary hash maps
(\texttt{elem\_position\_map} and \texttt{free\_list\_map} in Figure~\ref{fig:GraphStorageOptimizations}) for storing high-degree nodes on the PIM side.
The \texttt{elem\_position\_map} stores the mapping of edge to the edge's position in \texttt{cols\_vector}, and
the \texttt{free\_list\_map} stores the free positions in \texttt{col\_vector}.
For graph update, the host CPU only assumes simple tasks of writing data to a certain position within the \texttt{cols\_vector},
while the PIM side undertakes complex operations of edge retrieval and space management.
For example, \sysname{} follows this process to insert an edge <1, 2> in
Figure~\ref{fig:GraphStorageOptimizations}. 
Firstly, the \texttt{elem\_position\_map}
confirms that the edge does not exist. Then, the \texttt{free\_list\_map} allocates a
free space for the edge and the position is 1. Next, update the
\texttt{elem\_position\_map} by inserting (edge = <1, 2>, pos = 1). Finally, the
host CPU writes 2 to position 1 of the \texttt{cols\_vector} with row = 1.

Eventually, it forms heterogeneous graph storage for high-degree nodes, where
the CPU side enhances memory access locality while the PIM side overtakes the
majority of update expenses.


\section{evaluation}



\subsection{Evaluation Setup}

\textbf{Dataset.}
Our experiments use 15 real-world graphs from the SNAP\cite{snapdateset} dataset.
The 15 large-scale graphs (Trace ID \#1-\#15) with the number of nodes exceeding 200K are shown in Table~\ref{tab:dataset},
where nodes with out-degrees exceeding 16 are considered high-degree nodes.

\textbf{Baselines.}
We use RedisGraph\cite{RedisGraph} as a baseline system to evaluate our PIM system.
By representing the data as sparse matrices and employing highly optimized sparse matrix operations\cite{GraphBLAS},
RedisGraph delivers a fast and efficient way to store, manage, and process graphs\cite{redisGraphPerformance}.

We also implement the PIM-hash system as a contrast system, where all graph
nodes are distributed to PIM modules using a hash function, as the scheme of hash
partition is widely used in distributed graph databases\cite{chen2022g, li2022bytegraph}.

\textbf{Configurations.}
We use a server with two Intel(R) Xeon(R) Silver 4126 CPUs and 20 UPMEM DIMMs as the host system.
Each CPU has 16 cores at 2.10 GHz, and the L3 cache size is 22MB.
Each UPMEM DIMM has two ranks; each rank has 64 PIM modules.
RedisGraph utilizes a dedicated CPU core, exclusively benefiting from L3 cache and memory bandwidth.
\sysname{} and PIM-hash use a dedicated CPU core and 64 PIM modules (a rank).

\textbf{Workload Setup.}
For simplicity, our evaluation focuses on a typical RPQ, $k$-hop path query with a fixed start node.
The start node is randomly selected, and RPQs are processed in batch (batch size = 64K).
To show graph update performance, we randomly select 64K edges for insertion
and deletion.


\subsection{Performance of RPQs}

Figure~\ref{fig:kHopPathQuery} shows the run time of processing $k$-hop path
queries on \sysname{}, PIM-hash, and RedisGraph. Under the 15 real-world graphs
from the SNAP dataset, \sysname{} has the best performance. For the graphs with
less skewness (\#1, \#2 ,\#3, \#7, \#13, \#14, and \#15), \sysname{} outperforms
RedisGraph by 2.54-10.67x.
By dispatching path matching tasks to PIM modules and reducing data movement, 
\sysname{} breaks the "memory wall" bottleneck of path matching and achieves high performance.
When handling highly skewed graphs (\#5, \#6,
\#8, \#11, and \#12), \sysname{} outperforms the PIM-hash up to 2.98x. 
On the one hand, \sysname{} tackles graph skewness with the locality-aware node distribution, 
achieving load balance among PIM modules;
on the other hand, \sysname{} distributes skewed workloads to the host CPU, leveraging the strengths both
of the host side and the PIM side.

\begin{figure}[t]
  \centering
  \includegraphics[width=7.5cm]{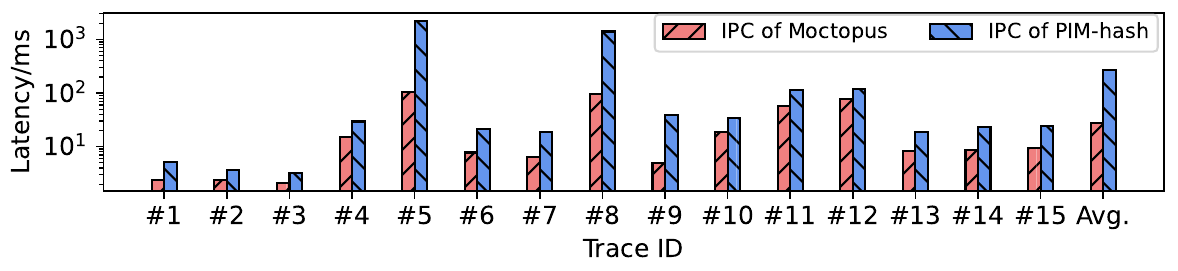}
  \vspace{-4.mm}
  \caption{IPC cost of \sysname{} and PIM-hash processing $3$-hop path queries (log scale).}
  \vspace{-4.mm}
  \label{fig:rpqCommunication}
\end{figure}

We further analyze the IPC cost during path matching.
Figure~\ref{fig:rpqCommunication} shows the IPC cost for \sysname{} and
PIM-hash processing 3-hop path queries. For $k$=3, \sysname{} reduces the IPC
cost by 89.56\% on average compared with PIM-hash, demonstrating that our graph
partitioning algorithm effectively preserves graph locality, ultimately
enabling more next-hops hit in local PIM modules.

It is observed that \sysname{}'s performance significantly deteriorates compared to RedisGraph as
$k$ increases. The reason is that, for most graphs except for road network graphs (\#1, \#2, and \#3), the number of
matched paths increases significantly with the increase of k, causing CPC and reduction to become the performance bottleneck.
For long path queries, we only perform $k$-hop path queries ($k$=4, 6, and 8) on road network graphs, and \sysname{} outperforms
RedisGraph by 6.00-9.71x.


\subsection{Performance of graph update}

\begin{figure}[t]
  \centering
  \includegraphics[width=7.5cm]{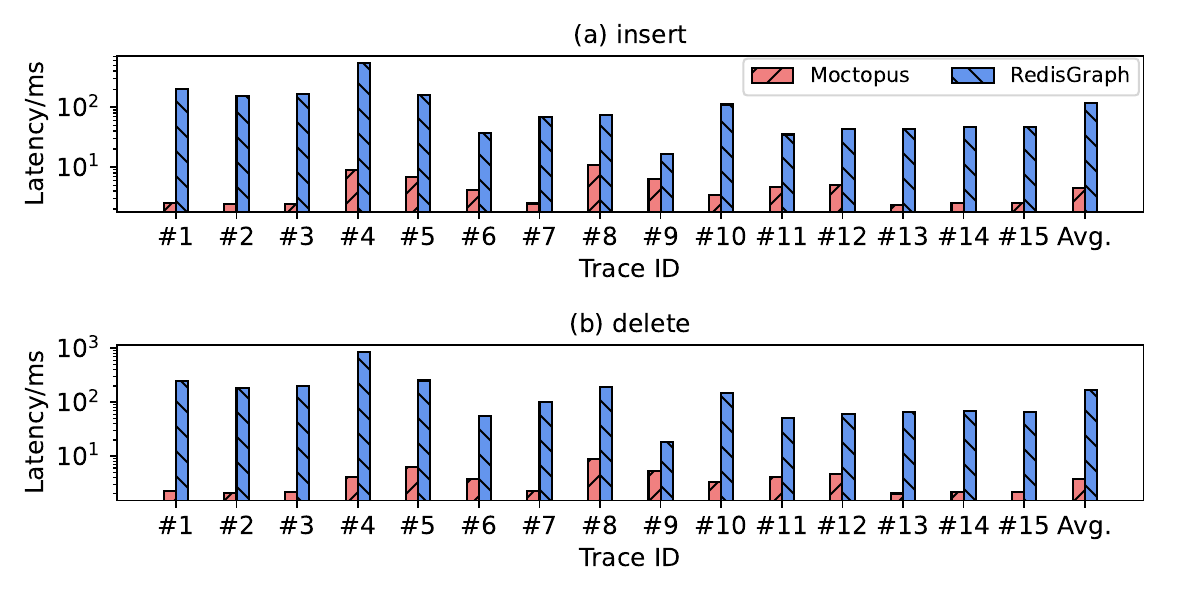}
  \vspace{-4.mm}
  \caption{Run-time of graph update (log scale).}
  \vspace{-6.mm}
  \label{fig:graphUpdate}
\end{figure}

Figure~\ref{fig:graphUpdate} shows the run-time of graph update (insert 64k
edges and delete 64k edges) on the 15 real-world graphs from the SNAP dataset.
Exempt from IPC and reduction stages, the graph update workloads can fully
utilize the intra-PIM bandwidth, resulting in exceptional performance. Compared
with RedisGraph, \sysname{} achieves up to 81.45x higher throughput with an
average of 30.01x for insertion and achieves up to 209.31x higher throughput
with an average of 52.59x for deletion. For the graphs with a high proportion
of graph data stored on the host side, the host CPU's load becomes a bottleneck
of graph update. Thanks to the heterogeneous graph storage for high-degree
nodes, \sysname{} amortizes the host side's update cost to the PIM side, still
achieving good graph update performance with an average spreading time of
50-160ns.




\section{Conclusion}

In this paper, we present \sysname{}, the first PIM system for accelerating
path matching over graph database with Processing-in-Memory. \sysname{} successfully
supports efficient batch RPQs and graph updates. The dynamic graph partitioning algorithm in
\sysname{} successfully tackles graph skewness and preserves graph locality
with low overhead. With PIM-friendly graph partitioning, \sysname{} addresses the
challenges of load imbalance and communication bottleneck during performing
RPQs. The optimizations for graph storage enables frequent graph update by
amortizing the host side's update cost to the PIM side. Evaluation against
RedisGraph and the widely-used hash scheme shows that \sysname{} is a PIM-friendly design and
can achieve high path matching and graph update performance.

\bibliographystyle{ACM-Reference-Format}
\bibliography{rpq-short}

\end{document}